\documentclass{Interspeech}

\interspeechcameraready

\title{WTFormer: A Wavelet Conformer Network for MIMO Speech Enhancement with Spatial Cues Peservation\thanks{$^{*}$ Corresponding author}}
\author[affiliation={1,2}]{Lu}{Han}
\author[affiliation={3}]{Junqi}{Zhao}
\author[affiliation={1,2,*}]{Renhua}{Peng}

\affiliation{Laboratory of Noise and Audio Research, Institute of Acoustics, Chinese Academy of Sciences}{Beijing}{China}
\affiliation{University of Chinese Academy of Sciences}{Beijing}{China}
\affiliation{Centre for Vision, Speech and Signal Processing (CVSSP), University of Surrey}{Guildford}{UK}
\email{\{hanlu2023, pengrenhua\}@mail.ioa.ac.cn, junqi.zhao@surrey.ac.uk}


\usepackage{comment}
\usepackage{graphicx} 

\begin{document}

\maketitle

\renewcommand{\thefootnote}{\fnsymbol{footnote}}
\footnotetext[1]{Corresponding author.}
\renewcommand{\thefootnote}{\arabic{footnote}}

\vspace{-0.3cm}
\begin{abstract}
Current multi-channel speech enhancement systems mainly adopt single-output architecture, which face significant challenges in preserving spatio-temporal signal integrity during multiple-input multiple-output (MIMO) processing. To address this limitation, we propose a novel neural network, termed WTFormer, for MIMO speech enhancement that leverages the multi-resolution characteristics of wavelet transform and multi-dimensional collaborative attention to effectively capture globally distributed spatial features, while using Conformer for time-frequency modeling. A multi task loss strategy accompanying MUSIC algorithm is further proposed for optimization training to protect spatial information to the greatest extent. Experimental results on the LibriSpeech dataset show that WTFormer can achieve comparable denoising performance to advanced systems while preserving more spatial information with only 0.98M parameters.
\end{abstract}

\section{Introduction}

Speech enhancement aims to recover clean target speech from noisy mixtures. Traditional single-channel methods \cite{boll1979suppression, fevotte2009nonnegative} relied on signal processing and statistical modeling.  However, these algorithms suffer great performance degradation in non stationary noise and low signal-noise ration (SNR) scenarios.
Multichannel beamforming algorithms focus on exploiting the spatial properties based on microphone arrays to suppress noise. Conventional beamforming, such as minimum variance distortionless response (MVDR) \cite{capon1969high} and generalized sidelobe cancelers (GSC) \cite{griffiths1982alternative}, enhance signals through adaptive beamforming and noise covariance matrix estimation. Nevertheless, their performance highly depends on accurate direction-of-arrival (DOA) estimation and assumptions regarding noise statistics.

The advent of deep learning has revolutionized speech enhancement techniques \cite{zheng2023sixty}. Single-channel deep neural networks can significantly improve speech quality and intelligibility by learning an end-to-end mapping from noisy to clean spectra or by predicting time-frequency (TF) masks. The integration of neural networks with multichannel signal processing techniques has resulted in the development of hybrid frameworks. These paradigms can be roughly classified into two main categories. The first category involves mask-based neural beamformers, which utilizes deep neural networks (DNNs) \cite{heymann2016neural, erdogan2016improved} to predict TF-masks, thereby improving covariance matrix estimation.
 Although these methods can improve the generalization ability of traditional algorithms, they are still limited by error propagation in cascaded systems. The second category includes end-to-end neural networks or neural spatio-spectral filters \cite{luo2019fasnet, tan2022neural} for implicit beamforming. This could theoretically allow for better performance, but it may also cause greater distortion.

Among these frameworks, the MIMO neural network model can suppress the unwanted noises while preserving spatial cues, making it well-suited for pre-processing. However, preserving spatial cues with neural networks is challenging due to the lack of a clear structural pattern in the phase spectrum \cite{zheng2018phase}. Nerual MIMO framework are often simply extended \cite{han2020real, kim2021mimo, halimeh2022complex, kimura2024diffusion} from single-channel configurations or used as the first stage \cite{ren2021causal, pandey2022tparn, wang2021multi} in Multi-Input Single-Output (MISO) model. Furthermore, current research on speech enhancement that preserves spatial information mainly focused on binaural aspects \cite{han2020real, tokala2024binaural}, with an emphasis on human auditory perception. In contrast, most beamforming algorithms are highly sensitive to phase, and rely on phase information for DOA estimation. Multi-channel speech enhancement algorithms employing microphone arrays face inherent challenges in maintaining spatial fidelity while achieving noise suppression, particularly in preserving interaural cues crucial for sound localization in MIMO systems.

To address these coexisting requirements of MIMO architectures, we propose WTFormer, a MIMO speech enhancement model combining wavelet convolution and Conformer. This approach leverages the multi-resolution properties of wavelet transform to expand the receptive field of the encoder's convolution kernel, enhancing the capture of global spatial features. During the temporal modeling phase, we employ TF-Conformer, which presents promising results in \cite{abdulatif2024cmgan}. Additionally, multi-dimensional collaborative attention (MCA) is utilized to better integrate time, frequency, and spatial information. A loss function based on the Multiple Signal Classification (MUSIC) algorithm, combined with a multi-task loss strategy, is designed to minimize spatial distortion during optimization. Evaluation on public dataset LibriSpeech, using a model with a minimal number of parameters, demonstrates comparable noise reduction performance to current multi-channel speech enhancement models \cite{li2022embedding}, while preserving more spatial cues.


\begin{figure*}[htbp]
\begin{center}
\includegraphics[width=0.8\textwidth]{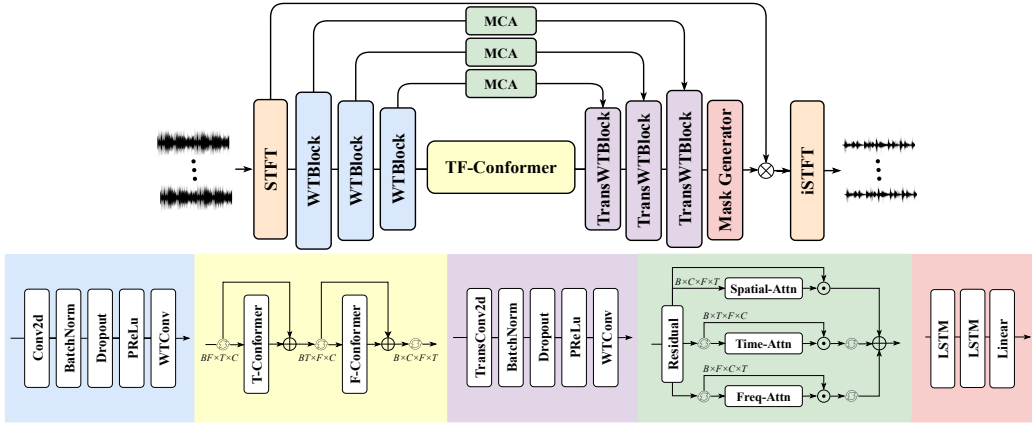}
\end{center}
\caption{An overview of the proposed WTFormer architecture.Different modules are remarked with different colors.} 
\label{fig1}
\end{figure*}

\section{Signal Model and Problem Formulation}
The signal recorded by a uniform linear M-channel microphone array can be expressed in the short-time Fourier transform (STFT) domain as: 
\begin{align}
\mathbf{Y}_{f,t}=\mathbf{S}_{f,t}+\mathbf{N}_{f,t}=H_{s}{S}_{f,t}+H_{n}{N}_{f,t},
\end{align}
where$\{\mathbf{Y}_{f,t},\mathbf{S}_{f,t},\mathbf{N}_{f,t}\}\in\mathbb{C}^M$ denotes the reverberant-noisy mixture speech, target speech and noise for $\mathbf{M}$ channels, with frequency index $f\in\{1,\cdots,F\}$ and time index $t\in\{1,\cdots,T\}$,  respectively. $H_{s}$ and $H_{n}$ denote multichannel relative transfer function (RTF) representing speech and noise. ${S}_{f,t}$ and ${N}_{f,t}$ represent speech and noise source signals. Using early-reverberation as learning target is better than using direct-path and dry clean signal for noise reduction \cite{wang2025systematic}, $H_{s}{S}_{f,t}$ can be further decomposed as: 
\begin{align}
H_{s}{S}_{f,t}=\mathbf{S}_{f,t}^{early}+\mathbf{S}_{f,t}^{late}=H_{s}^{early}{S}_{f,t}+H_{s}^{late}{S}_{f,t},
\end{align}
Where $\mathbf{S}_{f,t}^{early}$ is early-reverberation speech, $H_{s}^{early}$represents the direct and early responses of the Room Impulse Response (RIR), $H_{s}^{late}$ represents the late reverberation of the RIR. Early-reverberation speech is set as the target of model learning, and all other components are regarded as noise. The proposed approach estimates the clean speech as follow:
\begin{align}
    \hat{\mathbf{S}}_{t,f}^{early}\ = \mathcal{G}_{\Theta}(\mathbf{Y}_{t,f}),
\end{align}
where $\mathcal{G}_{\Theta}$ is the estimated MIMO complex masking filters by our system. In theory, it is possible to retain all spatial information and perform speech enhancement by learning a complex mask for each channel.

    

\section{Proposed WTFormer}

\subsection{System overview}

The model takes multi-channel input in the form of time-frequency domain representations of noisy speech signals. Initially, the time-domain signal is processed using the short-time Fourier transform (STFT) to extract time-frequency features. We adopt convolutional encoder-decoder (CED) structure with skip connections, which has been proven effective for speech enhancement. The encoder utilizes multiple layers of wavelet transform convolution block to progressively capture information in both the time and frequency domains. An intermediate TF-Conformer module is employed for time-frequency processing, combining the advantages of convolution and multi-head attention to efficiently extract and enhance features. Instead of traditional skip connections, multiple parallel multi-channel attention (MCA) modules are used. These spatial, temporal, and frequency attention modules (Spatial-Attn, Time-Attn, and Freq-Attn) work in parallel to enhance the representations compressed by the encoder, capturing both local and global dependencies in the feature space. This approach helps preserve important spatial information from the microphone array. Finally, a simple multi-layer recurrent neural network (RNN) predicts the mask and reconstructs the enhanced speech signal, followed by the application of inverse STFT to obtain the clean output.


\subsection{Wavelet Convolutions Block}
We introduce a wavelet transform-based waveform module (WTConv) \cite{finder2024wavelet} to enhance the spatial information retention capability and multi-frequency feature efficiency in multi-channel speech signal processing. WTConv uses Haar wavelet basis to perform multi-level stratification of the input signal to generate low-frequency approximation (LL) and high-frequency detail reduction (LH, HL, HH). This design not only captures long-term dependencies in speech signals but also enhances the transient noise characteristics by applying high-frequency gain, thereby improving the robustness of speech enhancement algorithms in complex acoustic environments. Additionally, the local temporal characteristics of wavelet transform preserve the spatio-temporal structure of speech signals and prevent phase distortion, which can occur in Fourier transforms during frequency domain operations, thus preserving the spatio-temporal consistency of multi-channel signals.

In the implementation, WTConv is integrated into the encoder-decoder layers of a MIMO speech enhancement network, alternating with traditional 2D convolution modules to expand the receptive field. Specifically, each WTBlock consists of a Conv2d layer, followed by a batch normalization layer, a dropout layer, a PReLU activation function, and a WTConv. The TransWTBlock mirrors the WTBlock structure, replacing the Conv2D with Transposed Conv2D.

\subsection{TF-Conformer Block}
Although the Conformer architecture \cite{gulati2020conformer} has demonstrated remarkable success in speech recognition, it remains underexplored for speech enhancement tasks. To effectively model both TF dependencies and spatial characteristics in multi-channel speech enhancement, we employ the TF-Conformer module, inspired by \cite{abdulatif2024cmgan}, as the intermediate processing module for embedding. 
The module sequentially processes time-frequency features through two cascaded conformer blocks, preserving spatial correlations across channels, and adaptively fuses the enhanced features with the original input via residual connections.
This dual-scale paradigm optimizes global temporal-frequency relationships and local spatial correlations, offering a robust solution for beamforming-free multi-channel speech enhancement systems.

Each Conformer module contains a half-step feed-forward networks (FFN), a multi-head self-attention (MHSA) mechanism block, a Conv-block and another FFN in sequence.
The Conv-block architecture comprises: 1) layer normalization followed by a gated linear unit (GLU)-activated point-wise convolution, 2) a swish-activated 1D depthwise convolution layer, and 3) a final point-wise convolution with dropout. All constituent sub-blocks incorporate residual connections to maintain gradient flow and preserve original signal fidelity.

\subsection{MCA Block}
In this paper, we proposes to use multidimensional collaborative attention module (MCA) \cite{yu2023mca} to replace the traditional skip connection structure. The MCA block utilizes a three-branch architecture that models attention across the spatial, time, and frequency dimensions in parallel, dynamically capturing global contextual dependencies of speech features. Specifically, each branch operates on different dimensions of the input features, combining global average and standard deviation pooling information through a Squeeze Transformation to generate an adaptive multidimensional feature descriptor. The Excitation Transformation then applies a lightweight local interaction mechanism to assign dynamic weights to features from different dimensions. Finally, the outputs of the three attention branches are averaged and aggregated, then fused with the original features via a sigmoid function to achieve cross-dimensional collaborative enhancement. This module strengthens the key time-frequency components and channel correlation of the speech signal with extremely low computational overhead, while alleviating the redundancy problem of information transmission in traditional skip connections.
\subsection{Mask Generator}
The Mask Generator is a crucial component of the proposed model, responsible for estimating the complex ideal ratio mask (cIRM) \cite{williamson2015complex} to enhance the noisy speech signals while preserving spatial information. The whole consists of two layers of long short-term memory network (LSTM) and one linear layer.

\begin{table*}[htbp]
\begin{center}
    \renewcommand{\thetable}{2}
    \caption{Results comparison with advanced baselines.}
    \begin{tabular}{@{}l|c|cccc|ccc@{}}
        \toprule
        Systems & Para.(M) & PESQ$\uparrow$ & STOI$\uparrow$ & eSTOI$\uparrow$ & SI-SNR(dB)$\uparrow$ & $\Delta$ITD($\mu$s) $\downarrow$ & $\Delta$IPD(rad)$\downarrow$  & $\Delta$ILD(dB)$\downarrow$ \\
        \hline
        noisy &- &1.64 & 0.67 & 0.45 & -0.97 &285.22 & 0.86 & 2.94\\
        Ti-MVDR & - &2.48 & 0.86& 0.73&7.54 & -& -&- \\
        MB-MVDR & - &2.53 & 0.88& 0.79&8.21 & -& -& -\\
        MIMO-UNet &1.96 & 2.14 & 0.80 & 0.71 & 6.78 &115.93 &0.81 & 0.89 \\
        EaBNet & 2.84&2.99 & \textbf{0.92} & \textbf{0.84} & \textbf{10.55} & 231.69& 0.85&1.43  \\
        WTFormer & 0.98&\textbf{3.02} &\textbf{0.92} &\textbf{0.84} &10.31 &\textbf{84.27} &\textbf{0.75} & \textbf{0.73} \\
        \bottomrule
    \end{tabular}
\end{center}
\end{table*}

\vspace{-0.2cm}

\section{Experiment}

\subsection{Dataset Preparation}
We used the public speech dataset LibriSpeech and multi-channel RIR to generate microphone-array signals for experiments. The uniform linear array (ULA) with 4 cm space interval and eight elements was used. The train-360 corpus was randomly split: 90\% for training, 5\% for verification, and 5\% for evaluation. The multi-channel RIR was generated 
using the image method. The room's length and width were set randomly between 5–10 m, and the height between 3–4 m. The microphone array center was randomly placed in the room, with at least 1 m from each boundary. Then, randomly rotations are added to the array in the x-y-z directions.


The speech source is placed 0.75–2 m from the array center, with at least 0.5 m from each wall. Noise source locations are generated similarly to the speech source. The SNR of training data ranges from -5 to 20 dB, while the test data ranges from -5 to 5 dB. The sampling rate is 16 kHz, the speed of sound is 343 m/s, reverberation time is 0.3–0.7 s, and data is chunked into 4 seconds segments. Additionally, the dynamic range of the audio is reduced within the range of [0.2, 0.9] for all data. A total of 335,735 training samples, 48,651 validation samples, and 48,652 test samples are obtained.

\subsection{Expereimental settings}
\subsubsection{Model Details}

In the three-layer WTBlock of the encoder, the kernel size of Conv2d is set as (6, 2) (7, 2) and (7, 2) with stride (2, 1) in the and frequency time axes. All dropout rate is 0.2, and WTConv has a kernel size of 5 with a stride of 1. In the Conformer Block, the Conv kernel size is 31 and Multi-head attention uses four heads. The pooling type of MCA block is selected as Average Pooling.

\subsubsection{Training Details}
All the utterances are sampled at 16 kHz. The Hanning window is utilized with 50\% overlap between adjacent frames and the frame length is set as 20 ms. After STFT, the real and imaginary parts are obtained, which are concatenated along the frequency dimension to obtain the time-frequency representation $\mathbf{Y}\in\mathbb{C}^{M\times 2F\times T}$, where $M$=8 is the number of array channels, $F$=161 is number of frequency bins, and $T$=401 is the frame number. All the models are trained with Adam optimizer \cite{kingma2014adam}, and the learning rate is initialized at 4e-4 and will be halved if the loss does not decrease for consecutive four epochs. The batch size is 16 and the number of epochs is 80. Automatic mixed precision training \cite{micikevicius2017mixed} is utilized for efficient training.

\subsection{Loss function}
From the perspective of loss function, we regard MIMO smpeech enhancement with spatial cues preservation as a multi-task learning task \cite{kendall2018multi}, optimizing both speech enhancement and spatial preservation. Speech enhancement is prioritized with a higher weight, and two learnable parameters, $\sigma_1$ and $\sigma_2$, dynamically adjust this weight. The overall formula is as follows:
\vspace{-0.35cm}
\begin{align}
\vspace{-0.5cm}
\mathcal{L}_{total}=\frac{10}{2\sigma_1^2}\mathcal{L}_{ns}+\frac{1}{2\sigma_1^2}\mathcal{L}_{ps}+\log(\sigma_1\sigma_2),
\vspace{-0.5cm}
\end{align}
where $\mathcal{L}{ns}$ represents the loss for noise suppression, and $\mathcal{L}{ps}$ is the loss for preserving spatial cues. For noise suppression, we use a loss function based on scale-invariant signal-to-noise ratio (SI-SNR). SI-SNR Loss \cite{isik2016single} has been shown to have good performance in speech enhancement tasks. For spatial information retention, we use the mean square error (MSE) of the MUSIC spatial spectrum of the multi-channel signal before and after processing as the loss function. MUSIC is used for estimating the DOA, which can effectively capture and decode spatial cues. For wideband speech, we divide it into 300 narrowband signals of frequency bands. After calculating a sample of 4 seconds, a spatial spectrum of 300×181 dimensions is obtained. Minimizing the MSE of the MUSIC spatial spectrum encourages the model to retain the spatial characteristics of the input signal. This loss function balances speech enhancement and spatial retention, ensuring the model preserves spatial cues while denoising.


\subsection{Baseline systems}

Four baseline approaches are selected: Ti-MVDR, MB-MVDR \cite{erdogan2016improved}, MIMO-UNet \cite{ren2021causal}, and EaBNet \cite{li2022embedding}. The first two methods are traditional MISO beamforming, assuming prior knowledge of target speech and noise, followed by beamforming weight calculation. The last two approaches are deep neural network based, where the filter-and-sum step is removed for MIMO comparison. Performance was evaluated using four metrics: PESQ \cite{rix2001perceptual}, STOI, eSTOI \cite{jensen2016algorithm}, and SI-SNR \cite{le2019sdr}, where higher values indicate better performance. The preservation of spatial information is evaluated using $\Delta$ITD, $\Delta$IPD, and $\Delta$ILD, commonly used in binaural audio tasks, where smaller values indicate better capability. We calculate the difference between microphones \{1, 5\} \{2, 6\} \{3, 7\} and \{4, 8\}, averaging the four groups after subtracting the target value for evaluation.

\section{Results and Discussion}

\subsection{Ablation Study}


We conduct the ablation study on WTFormer as shown in Table 1, where WTFormer-WT, WTFormer-MCA, WTFormer-$\mathcal{L}_{ps}$ indicate the removal of the WTConv, MCA, and $\mathcal{L}_{ps}$ loss, respectively. It can be seen that ablation of WTConv slightly degrades the $\Delta$ITD, but greatly affects the PESQ scores. This implies that WTConv improves the denoising performance through the large receptive field provided by its wavelet transform, while playing a critical role in preserving $\Delta$ITD information. It can also be seen that the ablation of MCA degrades both the $\Delta$ITD and PESQ, which confirms its irreplaceable role in capturing cross-channel dependencies. The $\mathcal{L}_{ps}$ loss function ablation results show a significant reduction in $\Delta$ILD and $\Delta$ITD, which validates the effectiveness of $\mathcal{L}_{ps}$ loss in preserving spatial cues.

\begin{table}[htbp]
\begin{center}
    \renewcommand{\thetable}{1}
    \caption{Ablation study results on the proposed WTFormer}
    \begin{tabular}{l|c|c|c}
        \hline
        Systems & PESQ$\uparrow$ & $\Delta$ ITD($\mu$s) $\downarrow$ &$\Delta$ ILD(dB)$\downarrow$\\
        \hline
        WTFormer-WT & 2.92 & 96.48 & 0.73 \\
        WTFormer-MCA & 2.95 & 102.15 & 0.75 \\
        WTFormer-$\mathcal{L}_{ps}$ & 3.02 & 104.39 & 0.82 \\
        WTFormer & 3.02 & 84.27 & 0.73 \\
        \hline
    \end{tabular}
\end{center}
\vspace{-0.8cm}
\end{table}

\subsection{Results Comparison with Advanced Baselines}
\begin{figure}[htb]   
\center{\includegraphics[width=8.3cm]  {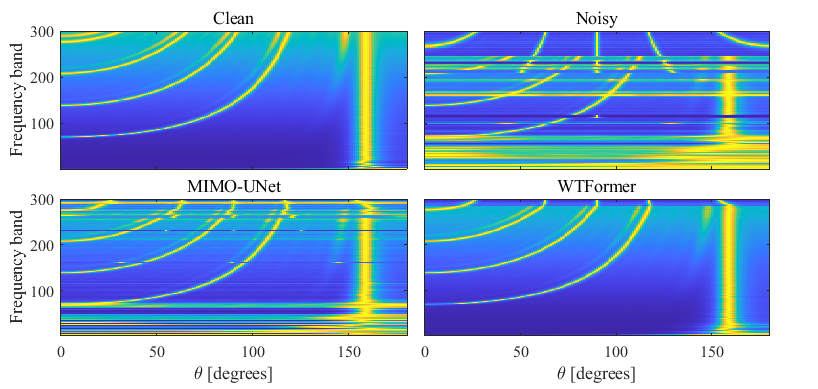}}   
\caption{\label{1} Spatial spectrum calculated by MUSIC algorithm.}   
\vspace{-0.5cm}
\end{figure}

Table 2 compares the proposed WTFormer with advanced baselines in terms of speech enhancement and spatial cues preservation metrics. The results demonstrate WTFormer's superiority in multiple aspects. WTFormer achieves the highest PESQ score among all systems and is comparable to EaBNet in terms of STOI (0.92) and eSTOI (0.84), with a slight SI-SNR degradation. This suggests that WTFormer may perform poorly in time-domain evaluation metrics, while being generally comparable to EaBNet in speech enhancement performance. It should be pointed out that WTFormer achieves this with only 0.98M parameters, which is significantly fewer than EaBNet and MIMO-UNet, demonstrating its high parameter efficiency.

WTFormer excels in spatial cues preservation, achieving optimal values for all related indicators. WTFormer reduces $\Delta$ITD by 27.3\% and $\Delta$ILD by 18.0\% compared to MIMO-UNet. This significant improvement highlights WTFormer's ability to preserve spatial information while enhancing speech quality. It can be seen that MIMO-UNet can achieve better spatial cues retention than that of EaBNet at the expense of PESQ score degradation. In contrast, our hybrid network WTFormer, achieves the best spatial retention without sacrificing noise reduction by expanding the receptive field and using multi-dimensional collaborative attention. Figure 2 shows the spatial spectrum calculated using the MUSIC algorithm. Severe interference in both high- and low-frequency bands compromises the accuracy of DOA estimation in noisy signals. The spatial cues recovered by MIMO-UNet is partially restored, but remains blurred in many low-frequency parts. The spatial spectrum of WTFormer closely matches the clean speech, indicating excellent preservation of spatial information. Only slight distortion appears in the highest frequency band. This may be due to the few speech components in this frequency band, allowing noise to dominate.


\section{Conclusions}

This paper introduces WTFormer, a novel MIMO speech enhancement framework that preserves spatial cues while achieving competitive noise reduction. By integrating wavelet convolutions for multi-resolution analysis, TF-Conformer blocks for time-frequency modeling, and multidimensional collaborative attention for spatial dependency learning, WTFormer maintains essential inter-channel phase and magnitude relationships. The use of a MUSIC-based spectral loss further enhances spatial fidelity. Experimental results show that WTFormer achieves superior denoising with 0.98M parameters, outperforming baselines in spatial cues preservation.
Future work will explore more interpretable causal models.



\bibliographystyle{IEEEtran}
\bibliography{mybib}

\end{document}